# The fallacy of evidence based policy

Draft, Submitted for a special issue
on FUTURES, January 2015

Andrea Saltelli (1,3), Mario Giampietro (2,3)
(1) Centre for the Study of the Sciences and the Humanities (SVT),
University of Bergen (UIB)
(2) Institució Catalana de Recerca i Estudis Avançats  (ICREA)
(3) Institut de Ciència i Tecnologia Ambientals (ICTA),
Universitat Autonoma de Barcelona (UAB)

**Abstract**

The use of science for policy is at the core of a perfect storm generated by the insurgence of several concurrent crises: of science, of trust, of sustainability. The prevailing modern positivistic model of science for policy, known as 'evidence based policy', is based on dramatic simplifications and compressions of available perceptions of the state of affairs and possible explanations (hypocognition). Therefore this model can result in flawed prescriptions. The flaws become more evident when dealing with complex issues characterized by concomitant uncertainties in the normative, descriptive and ethical domains.  In this situation evidence-based policy may concur to the fragility of the social system.

Science plays an important role in reducing the feeling of vulnerability of humans by projecting a promise of protection against uncertainties. In many applications quantitative science is used to remove uncertainty by transforming it into probability, so that mathematical modelling can play the ritual role of haruspices. This epistemic governance arrangement is today in crisis.

The primacy of science to adjudicate political issues must pass through an assessment of the level of maturity and effectiveness of the various disciplines deployed.  The solution implies abandoning dreams of prediction, control and optimization obtained by relying on a limited set of simplified narratives to define the problem and moving instead to an open exploration of a broader set of plausible and relevant stories.

Evidence based policy has to be replaced by robust policy, where robustness is tested with respect to feasibility (compatibility with processes outside human control); viability (compatibility with processes under human control, in relation to both the economic and technical dimensions), and desirability domain (compatibility with a plurality of normative considerations relevant to a plurality of actors).

**Key words:** Evidence based policy, science for governance, STS, Post-Nomal Science

# Section 1 – Science for policy. Predicaments and doubts

*The incoming commission must find better ways of separating evidence-gathering processes from the 'political imperative'.*

In a critique of science's use for policy A. Glover, former Chief Science Adviser of the President of the European Commission, laments that too often evidence based policy turns into its opposite, policy based evidence, and for this reason she appeals to a more rigid separation of science and policy (Wildson, 2014).

Are these diagnosis and therapy right? In this section we claim that this is not the case, and that the problem runs much deeper, where concomitant crises of science, trust and of sustainability call for different medicines that just separating evidence from policy.

We shall use mostly insights from two field of scholarship: the first is known as science and technology studies (STS), and the second is bioeconomics.

As per the former field we shall make frequent reference to style of policy engaged science described by Funtowicz and Ravetz as Post Normal Science (PNS) were "facts are uncertain, values in dispute, stakes high and decisions urgent" (Funtowicz and Ravetz, 1991, 1992, 1993). We deal here with epistemic governance, defined as "*how knowledge for policymaking should be understood and governed*" (Pearce and Raman, 2014).

As per the second field of inquiry we shall mostly follow on the tradition of Nicholas Georgescu Roegen and Robert Rosen (Rosen, 1991; Giampietro, 2013), where the epistemological implications of complexity are addressed at the moment of generating quantitative analysis.

## 1.1 Times ripe with controversy

It would be naïve to reconsider the place and the style of science support to policy neglecting the increasing climate of controversy which in recent years has marked practically all instances in which science has been used to a policy end. From the impact of pesticides on bees to the culling of badgers, from the greenhouse potential of the refrigerant liquid used by Mercedes Benz to the impact of endocrine disruptors, from the benefits of shale gas fracking to the fate of children raised by gay parents, from the true long term cost of citizenship for illegal migrants to the desirability of international testing and comparison of the educational attainment of children, all is matter of contention where the relevant science is disputed. Until recently these levels of antagonism were reserved for issues such as anthropogenic climate change and genetically modified organisms (GMO), the quintessential 'wicked' issues (Rittel & Webber 1973) - issues so deeply entangled in a web of hardly separable facts, interests and values that there cannot be agreements among different parties about the nature of the problem. Dan Kahan (2014) has observed that climate change affects us so intimately that it can define who we are culturally and normatively. We postulate that this kind of affection, and the related attitude toward scientific facts, can be found in many other issues, beside climate change, and that therefore, the unavoidable presence of cultural and normative bias cannot be silenced by scientific proficiency.

## 1.2 Science's own crisis and its roots

Science own house seems at present in a state of crisis. The Economist – a periodical - titles on his cover "How Science goes wrong" (2013). Its main editorial reads:

> *Science still commands enormous—if sometimes bemused—respect. But its privileged status is founded on the capacity to be right most of the time and to correct its mistakes when it gets things wrong.* […] *The false trails laid down by shoddy research are an unforgivable barrier to understanding.*

The Economist's piece draws from and an increasing number of academic papers and editorials lamenting a drop in reproducibility and a corresponding increase in retraction of scientific work. 'Unreliability in scientific literature' and 'systematic bias in research' are lamented by Boyd (2013, Nature). "Laboratory experiments cannot be trusted without verification", argues Sanderson for organic chemistry research (2013, *Nature*). 'Suspected work […in] the majority of preclinical cancer papers in top tier journals' is denounced by Begley (2013, *Nature*). In a landmark study of results in cancer science Begley and Ellis were able to reproduce only 11 per cent of the original findings (2012). That this may result in a death sentence for patients on experimental trials (with pharma having passed the pre-clinical phase) gives an idea of the seriousness of the issue.

The issue is not confined to natural and medical sciences. "I see a train wreck looming" warns Nobel laureate Daniel Kahneman with behavioral sciences in his sight (Yong, 2012); Joseph Stiglitz (2010), another laureate, condemns perverse incentives in the modelling of financial products at the hearth of the present economic recession.

According to The Economist's issue just quoted the main problem is a general lack of statistical skills among scientists, e.g. in balancing false positives and false negatives. Poor refereeing is also pointed to. The issue is perhaps more serious than can be corrected by statistical training. According to Ioannides (2005):

> *A research finding is less likely to be true when . . . there is greater financial and other interest and prejudice; and when more teams are involved in a scientific field in chase of statistical significance.*

The situation is so serious that a Meta-Research Innovation Centre (METRICS) has been launched at Stanford, involving the same John Ioannidis, to combat 'bad science' (The Economist, (2014). The same author contends (2014) that as a result of shoddy science as much as 85% of research funding is wasted. The peer review system at the hearth of science's quality control mechanism is itself in a status of crisis: suffices here to mention titles such as "Springer and Université Joseph Fourier release SciDetect to discover fake scientific papers" (Springer, 2015) and "China's Publication Bazaar" – and subtitled: "A Science investigation has uncovered a smorgasbord of questionable practices including paying for author's slots on papers written by other scientists and buying papers from online brokers".  See Funtowicz, and Ravetz, (2015) for a discussion.

The present crisis of science was predicted by Jerome R. Ravetz already in 1971. In his book 'Scientific Knowledge and its Social Problems' he notes (p.22):

> *[…] with the industrialization of science, certain changes have occurred which weaken the operation of the traditional mechanism of quality control and direction at the highest level. […]The problem of quality control in science is thus at the centre of the social problems of the industrialized science of the present period. If it fails to resolve this problem […] then the immediate consequences for morale and recruitment will be serious; and those for the survival of science itself, grave.*

The centrality of ethics for the quality and the self-governance of science so clearly illustrated by Ravetz is also central to the work of Jean-François Lyotard. In his 1979 work 'La Condition postmoderne. Rapport sur le savoir' he tackles the de-legitimization of knowledge (identified with science) when this becomes an industrialized commodity – as opposed to the instrument of emancipation and betterment of human beings (*bildung*). Philip Mirowski describes the degeneration of industrialized science in the US with painstaking detail in his work (2011) 'Science-Mart: Privatizing American Science'. According to Mirowski after the eighties neoliberal ideologies succeeded in decreasing state intervention in the funding of science, which became increasingly privatized and sub contracted, generating the perverse system of incentive already mentioned by Ioannides and Stiglitz.

In conclusion although we do not know whether the collapse in trust in science for policy is motivated by the problems of science for science, we suspect that we cannot tackle the latter without solving the former.

According to Lyotard (ibidem) it is since the age of Plato the question of the legitimacy of science is inextricably connected to that of the legitimacy of the legislator:

> '*Who decides what counts as knowledge and who knows about what one must decide?* […] *The question of knowledge in the information society is more than ever the question of government*'.

'*Solutions to the problem of knowledge are solutions to the problem of social order*' is a conclusion of Steven Shapin & Simon Schaffer 1985 book 'Leviathan and the Air-Pump: Hobbes, Boyle, and the Experimental Life'. All this implies that our chosen epistemologies are not issues for academicians and philosophers, but a direct concern for the entire polity.

Going back to Anne Glover's diagnosis and therapy we can now observe that a better separation between science and policy corresponds to a well identified epistemology: one known as 'demarcation model' (Funtowicz, 2006). This models aims at protecting science from the political interference, preventing possible abuse of science and scientific information driven by agendas. It prescribes a clear demarcation between the institutions (and individuals) who provide the science and those where it is used.

The demarcation model is challenged in more recent epistemologies, mostly based on the impossibility to achieve separation between facts and values. The idea of an algid and disinterested science which manages to interface itself cleanly with the messiness of the political process is an abstraction devoid of salience. Instead of purified facts we mostly deals with hybrid arrangements (Latour, 1991), and one of the features of the present epistemic governance crisis is that "*the more knowledge is produced in hybrid arrangements, the more the protagonists will insist on the integrity, even veracity of their findings*" (Grundmann, 2009).

Alternatives epistemologies are offered by 'Post Normal Science' – already mentioned - and by the 'Co-production of knowledge' (Jasanoff, 1996); a review is in Carrozza, (2014).

The attention here is to issues of participation, legitimacy, transparency and accountability. In the 'extended participation model' (Funtowicz, 2006) the deliberation is extended both across disciplines – acknowledging that different disciplines see though different lenses, and across communities of both experts and stakeholders. When adopting this model one moves from 'speaking truth to power' towards 'working deliberatively within imperfections' (Funtowicz and Ravetz, 1990, 1993; van der Sluijs et al., 2008; see also www.nusap.net).

The remark has been made that the movement known as 'Citizens' Science' should be more responsive to official science's predicaments (McQuillan, 2014) and 'pick up the gauntlet' thrown by official science's contested hegemony. PNS is suggested by these authors as a promising avenue to move in this direction. The same fathers of PNS ask speculatively "Is the internet to science what the Gutenberg press was to the church?" noting that "[…] the new social media have given strength to the extended peer community in science in a way reminiscent of the contribution of printing to the Reformation" and "Collegial peer review is being rapidly transformed to review by an

'extended peer community,' raising important issues to the governance of science"
(Funtowicz and Ravetz, 2015).

### 1.3 Trust, modelling, uncertainties

In a recent speech before the European Parliament Pope Francesco has recently sent a blunt 'the emperor has no clothes' message. A quote:

> *The great ideas which once inspired Europe seem to have lost their attraction, only to be replaced by the bureaucratic technicalities of its institutions. As the European Union has expanded, there has been growing mistrust on the part of citizens towards institutions considered to be aloof, engaged in laying down rules perceived as insensitive to individual peoples, if not downright harmful*

The link of this warning to science for policy may not seem obvious but it is there nonetheless, as rules are the result of policies, and policies are defended on the basis of 'evidence'. As a relevant example consider how science was recruited to advocate austerity in public budgets. A 90% ratio of public debt to gross domestic product was stipulated by Harvard professors Kenneth Rogoff and Carmen Reinhart as an absolute ceiling above which growth would be hampered. Thus debt ratios above this limit were defined as unsafe for a country. A later reanalysis by researchers from the University of Massachusetts at Amherst disproved this finding by tracing it to a coding error in the authors' original work. Clearly once this particular result was repudiated the policies had already be put in place and "*In Britain and Europe, great damage has been done as a result.*" (Cassidy, 2013).

This is but one of the many instances where improper use of mathematical modelling has been instrumental of supporting flawed policies. Modelling hubris and its consequences are discussed in (Saltelli et al., 2013, Saltelli & Funtowicz, 2014). In his 2013 work 'Never Let a Serious Crisis Go to Waste: How Neoliberalism Survived the Financial Meltdown' Philip Mirowski devotes a long section (pp 275-286) to the story of how dynamic stochastic general models (DSGE) were the subject of a hearing in the US senate – '*an event in 2010 that was literally unprecedented in the history of economic thought in America*', p. 275, with sworn testimony of economists such as Sidney Winter, Scott Page, Robert Solow, David Colander and V.V. Chari, to understand how 'theorists tools' had come to be used as policy instruments and why these instruments were all but useless in anticipating the economic crisis. Queen Elisabeth had a similar moment with British economists at the London School of Economics (Pierce, 2008).

Saltelli and Funtowicz (2014) list several problems in the way mathematical modelling is used to tame uncertainty in relation to the production of evidence for policy. These include the rhetorical or ritual use of possibly disproportionate mathematical models to impress or obfuscate, the reliance on tacit possibly unverified assumptions, the instrumental inflation or deflation of uncertainties according to expedience, the instrumental compression and linearization of the analysis as to tame complexity and to convey an impression of prediction and control, and finally an absent or perfunctory sensitivity analysis.

An original insight on science's lost hegemony comes from Dan Sarewitz (2000). For this scholar the problem is not a lack of Science but of its abundance:

> *"Rather than resolving political debate, science often becomes ammunition in partisan squabbling, mobilized selectively by contending sides to bolster their positions. Because science is highly valued as a source of reliable information, disputants look to science to help legitimate their interests. In such cases, the scientific experts on each side of the controversy effectively cancel each other out, and the more powerful political or economic interests prevail, just as they would have without the science."*

## Section 2. Vulnerability, uncertainty and governance.

### 2.1 Science to tame human vulnerability: the Cartesian Dream

Humans are reflexive anticipatory systems capable to monitor and predict, to a certain extent, events associated with their interaction with the environment (Rosen, 1985). This provided them with a major comparative advantage against other species. The domain of *homo sapiens* on the planet seems so pervasive that the present era has been termed Anthropocene. When taken as isolated individuals human beings are vulnerable and, as a species, humans still depend on processes outside their control for their surviving. These sources of vulnerability have forced humans to build their identity around organized communities and to adopt religious beliefs. The legitimization of power in a social group is based on the willingness of individuals to trade part of their autonomy, via social contracts, in exchange of a reduction of their vulnerability to a host of needs (in the sense of Maslow), such as personal security against hostile action of other human beings; food, energy and water security; housing and job security; health care; environmental security; cultural identity and participation. No human society was ever able to provide protection against death.   However, a reduction in the feeling of vulnerability to this event was possible through religion. In particular, introducing the concept of an eternal life after death attenuated the fear associated with the loss of the physical body, thus providing a considerable power to those religions which were able to sustain the claim.

For this reason, for a long period of time societies were ruled by a combination of religious and military power. With the French revolution and the onset of modernity the new ruling class replaced religion with science, following on the prophecies of Francis Bacon (1561-1626) and the more recent formulations of René Descartes (1596-1650) and Nicolas de Caritat, marquis de Condorcet (1743-1794,) on how all social problems would ultimately be tamed by knowledge. In modern times the prophecy's main tenets have been cogently expressed by Vannevar Bush (1890-1974) right at the end of War World II. In the synthesis of Rommetveit et al., 2013:

> *Bacon formulated the basic belief that knowledge gives us power to act in the world to the benefit of our lives. Condorcet elaborated the utopia of a science-based society as one of welfare, equality, justice and happiness. Bush argued that scientific progress and a strong public funding of basic science are necessary conditions to sustain economic growth by the development of new products (or innovation in contemporary vocabulary).*

Bacon's utopia, as described in the Magnalia Naturae, an appendix to the New Atlantis, includes "*wonders of nature, in particular with respect to human use*", which reads (Bacon, 1627):

> *The prolongation of life; The restitution of youth in some degree; The retardation of age; The curing of diseases counted incurable; The mitigation of pain; More easy and less loathsome purgings; The increasing of strength and activity; The increasing of ability to suffer torture or pain; The altering of complexions, and fatness and leanness; The altering of statures; The altering of features; The increasing and exalting of the intellectual parts; Versions of bodies into other bodies; Making of new species; Transplanting of one species into another; Instruments of destruction, as of war and poison; Exhilaration of the spirits, and putting them in good disposition; Force of the imagination, either upon another body, or upon the body itself; Acceleration of time in maturations; Acceleration of time in clarifications; Acceleration of putrefaction; Acceleration of decoction; Acceleration of germination; Making rich composts for the earth; Impressions of the air, and raising of tempests; Great alteration; as in induration, emollition, &c; Turning crude and watery substances into oily and unctuous substances; Drawing of new foods out of substances not now in use; Making new threads for apparel ; and new stuffs, such as paper, glass, &c; Natural divinations; Deceptions of the senses; Greater pleasures of the senses; Artificial minerals and cements.*

One century later  Condorcet was so convinced of physics' ability to solve human predicaments that in the Ninth Epoch of his 'Sketch for a Historical Picture of the Progress of the Human Spirit' he states: "*All the errors in politics and in morals are founded upon philosophical mistakes, which, themselves, are connected with physical errors*" (Condorcet, 1785).

Closer to our times Vannevar Bush's dream was couched in the Endless Frontier metaphor (1945):

> One of our hopes is that after the war there will be full employment. [...]To create more jobs we must make new and better and cheaper products [...] new products and processes are not born full-grown. They are founded on new principles and new conceptions which in turn result from basic scientific research. Basic scientific research is scientific capital [...] It has been basic United States policy that Government should foster the opening of new frontiers. It opened the seas to clipper ships and furnished land for pioneers. Although these frontiers have more or less disappeared, the frontier of science remains.

The reason we quote these very old or just old texts is that in spite of the copious ink poured by social sciences and STS scholarship in warning about the limits of Cartesian Dream, from Stephen Toulmin's 'Return to Reason' and 'Cosmopolis' to Paul Feyerabend's 'Against Method', from Lyotard's 'The Post-Modern Condition' to Bruno Latour's 'We have never been modern', and the fact that a major science war was fought between natural and human sciences in between the eighties and the nineties[i], the Cartesian dream is still a prevailing narrative. Jerome Ravetz (1971, p. 387) calls this the folk-science of the educated classes:

> Indeed, we may say that the basic folk-science of the educated sections of the advanced societies is 'Science' itself in various senses derived from the seventeenth-century revolution in philosophy. This is quite explicit in figures of the Enlightenment such as Condorcet [...] a basic faith in the methods and results of the successful natural sciences, as the means to the solution of the deepest practical problems [...]

.

As discussed by Stephen Toulmin in 'Cosmopolis, The Hidden Agenda of Modernity' the XVII century vision of Cosmopolis, a society as rationally ordered as the Newtonian physics, perpetrated - thanks to its extraordinary success in many fields of endeavour - an agenda of prediction and control; an agenda whereby ecosystems and social systems could be fitted into precise and manageable rational categories. The agenda now more than ever crashes against the complexities of the present crisis, endangering the legitimacy of the existing social contracts.

## 2.2 Reductionism, hypocognition and socially constructed ignorance

We started this paper with the 'evidence based policy model'. In this section we try to show that this model is based on a series of dramatic simplification and linearization which are congruent with the both the Cartesian dream and what has been termed 'Hypocognition' (Lakoff, 2010), 'Socially constructed ignorance' (Rayner, 2012), and much earlier by Jerome R. Ravetz, 'Usable Ignorance' (1987).

Confronted with complex issues the application of Cartesian method of reduction with the goal of individuating linear and direct causal explanations becomes inadequate. Within this paradigm the scientific advisors determining fish quotas will be brought to court either if they have set the quota too high (for the consequent collapse of the catch) or too low (for the consequent loss of income for the fishermen). Volcanologists can be sentenced to jail, as in the case Aquila's (Italy) earthquake - though the sentence was reversed on appeal - because of miscommunication on the unavoidable uncertainty associated with earthquake predictions.

Quantitative analysis is predicated on a selection of a problem structuring (the adoption of a frame). This determines in any case a major compression of the information space that is later on used for governance. The compression is operated both at the normative level (one world view is adopted – the choice of the WHY) and at the level of the representation - what are the salient attributes relevant for the description of the system – the choice of the HOW (Giampietro et al. 2013).

This process is explained by Rayner (2012) in terms of socially constructed ignorance, which is not the result of a conspiracy but of the sense-making process of individuals and institutions:

> *To make sense of the complexity of the world so that they can act, individuals and institutions need to develop simplified, self-consistent versions of that world. The process of doing so means that much of what is known about the world needs to be excluded from those versions, and in particular that knowledge which is in tension or outright contradiction with those versions must be expunged. […] But how do we deal with […] dysfunctional cases of uncomfortable knowledge […]?*

This compression comes to a cost and can lead to the degeneration of a given arrangement, when generalized and institutionalized, eventually producing a situation of Ancien Régime. Then the inability of the system to cope with stressors leads to a strategy of denial, and to the refusal to process either internal or external signals, including those of danger (Funtowicz and Ravetz, 1994).

The compression of the information space results into ignoring knowledge which is available in established scientific disciplines which is not considered in the given problem structuring. Rayner call these the "unknown knowns", e.g. that knowledge which exists out there in academia and society but is actively removed by the compression. For Rayner "*unknown knowns* [are] *those which societies or institutions* *actively exclude because they threaten to undermine key organizational arrangements or the ability of institutions to pursue their goals.*"

Also ignored after the compression are the "known unknowns" – knowledge of gaps and areas of ignorance which is available but is not considered as relevant in the chosen issue definition.

The result of this compression is to focus the attention of the analysts on a finite set of attributes and goals. This fatally calls for process of optimization, e.g. the analyst ends up investing time and energies to find the best solution in the wrong problem space.

Needless to say the hubris generated in this way increases fragility, foremost in relation to "unknown unknowns", as the optimization implies a reduction of the diversity of behaviours (because of the elimination of the less performing alternatives within the chosen problem structuring) and therefore a reduction of adaptability (because of the neglect of attributes and goals not considered in the optimization). The issue is discussed at length in Nassim N. Taleb work 'Antifragile' (2012).

A lesson from bioeconomics (Giampietro et al., 2013) is that science should be able to address and integrate relevant events and processes that can only be observed and described by adopting simultaneously non-equivalent narratives (dimensions of analysis) and different scales (descriptive domains). In this case the virtue of reductionism (making possible rational choices based on a clear identification of relevant attributes, goals and direct explanations) becomes a vice. A rationality based on a simple problem structuring that is applied to solve a complex issue becomes a "mad rationality" – a concept attributed to social philosopher Lewis Mumford . The example of bioethanol from corn, where hundreds of billions of tax payer money have been invested in developing an alternative energy source that consumes more or less the same amount of energy carriers that it produces can be a good example of this effect (Giampietro and Mayumi, 2009). The story could continue to discuss how the same biofuels are considered in the public discourse as a strategy to mitigate emissions, when this appear to be the case only because of cuts in food production (Searchinger et al., 2015).

Socially constructed ignorance can also be defined as the institutional hegemonization of a given story-telling - i.e. the pre-analytical choice of a given set of relevant narratives, plausible explanations and pertinent perceptions and representations - which is assumed, by default, to be valid in normative, descriptive and ethical terms to deal with any problem.

This choice may determine situations in which the elephant in the room goes unnoticed, especially after the chosen story-telling has been dressed by a convenient suite of selected indicators and mathematical modelling.

Famous instances of missed elephants are the presidential address to the American Economic Association of the Nobel laureate in Economics Robert Lucas in 2003 announcing that the "central problem of depression-prevention has been solved" once and for all; and the 2004 the 'great moderation' speech of Barnenke, Chair of the US Federal Reserve, about the successful taming of volatility of business cycle fluctuations. In both cases top ranking exponents of the ruling paradigm were unaware of the possibility of the financial collapse that would lead to the world economic crisis in the next years.

These blunders have fed into the long standing dispute for a reconsideration of the prevailing paradigm in Economics (Reinert, 2008, Mirowki, 2013), and even within the ranks of the discipline new curricula are being studied (INET, 2013). More radically some voices have called for a reconsideration of Economics as the authoritative discipline to adjudicate social and environmental issues (Ravetz, 1994, Giampietro, 2012, p.xx, Fourcade et al., 2014). Noting the state of the economic discipline as used to solve socioeconomic problems one cannot help considering the possibility that the discipline might have reverted to (or never emancipated from) a state of immaturity. In a chapter entitled 'Immature and ineffective fields of inquiry' Jerome R. Ravetz remarks (1971, p. 366):

> […] *The situation becomes worse when an immature or ineffective field is enlisted in the work of resolution of some practical problem. In such an uncontrolled and perhaps uncontrollable context, where facts are few and political passions many, the relevant immature field functions to a great extent as a 'folk-science'. This is a body of accepted knowledge whose function is not to provide the basis for further advance, but to offer comfort and reassurance to some body of believers.*

In order to prove the falsehood of inference constructed by heavy modelling weaponry one does not need the language of mathematics, but plain English. To make an example, a critique of the already mentioned dynamic stochastic general equilibrium models DSGE (used as policy instruments) is possible by falsifying the underlying hypotheses of 'efficient markets' and 'representative agent' (Mirowski, 2013, pp 275-286). This is why we keep referring to semantics and story-telling in relation to the issue of framing. There is nothing new in this approach. Translating into English the result of mathematical elaboration was a teaching of Alfred Marshall (Pigou, Ed., 1925, p. 427), a teaching which is not unknown to present day economists (Krugman, 2009, p. 9) but which is often neglected when using mathematical modelling as Latin, to obfuscate rather than to illuminate (Saltelli et al., 2013).

Note that we are discussing here the use of mathematical models as an input to policy, e.g. as a tool to generate inferences for policy. For Joseph Stiglitz (2011):

> *Models by their nature are like blinders. In leaving out certain things, they focus our attention on other things. They provide a frame through which we see the world.*

There is nothing wrong in using blinders in the quest for theoretical progress. The issue is when the same tool is used to prescribe policy, expediently neglecting the blinding stage. This is what Nassim Taleb (2007) condemns as an attempt to 'Platonify reality'. This is also one of the processes by which hypocognition is generated. Rayner see this as one of the strategies to socially construct ignorance and calls it 'displacement':

> *[…] displacement occurs when an organization engages with an issue, but substitutes management of a representation of a problem (such as a computer model) for management of the represented object or activity.*

Displacement does not imply wrong models – which could possibly be corrected, but irrelevant models, which cannot be corrected through "learning by doing" and hence can do damage for a longer period of time. Along similar lines Financial Times columnist Samuel Brittain notes (2011):

> *Nothing has done more to discredit serious economic analysis than its identification with the guesses about output, employment, prices and so on which politicians feel obliged to make. […] True scientific predictions are conditional. They assert that certain changes […] will, granted other conditions are met, […], lead to a certain state of affairs […]. But they cannot tell us that the required conditions will be fulfilled.*

Evidence based policy has thus reached a situation of paradox, where all know and repeat that a certain practice – displacement in Rayner's lingo – is incorrect, but it is pursued nevertheless. In this way society is led to associate the stabilization of its own wellbeing with the stabilization of the institutional settings determining the status quo.

## 2.3  Legitimacy versus simplification

So far the conventional scientific approach has dealt with sustainability issues trying to individuate the best course of action by using deterministic models.  This strategy assumes that it is possible to predict the behaviour of complex self-organizing systems (including those that are reflexive such as human societies) and that the quality of the scientific input to the policy process is ensured by the rigour of the methods deployed. This assumption overlooks the accumulation of uncertainties which – when properly appraised – implies the total inability of these tools to generate useful inference.

Thus we expect for example that modelling approaches which have failed to predict a purely financial and economic crisis will be able to inform us about the behaviour of a system involving institutions, societies, economies and ecologies, such as we do when applying the craft of cost benefit analysis (CBA) to climate change, and pretend to assess the impact on the economy of increased crime rates resulting from hotter temperatures (Rhodium Group, 2014).

Climate instead calls into operation the multiple connections involved in the nexus between energy, water, food and the reproduction of human institutions. The complexity of the organization of socio-ecosystems is not an issue to be solved, but an inherent property of the self-organizing systems considered, i.e. ecosystems and human societies. These systems express agency (reproducing themselves and interacting with each other) across a variety of relevant scales. When dealing with them analysing a dimension at the time (water, energy, food, land use) and a scale at the time (local, meso, macro) does not work and provides unlimited scope for undesired blows from the law of unintended consequences (Giampietro, 2013). When science is used to suppress uncertainty - rather than to explore the sources of our ignorance - failures are likely. These may take the form of adopting by default the "more of the same" policy, even if the specific policy did not work in the past. Scientific activity that is forced to operate systematically outside its field of applicability is likely to become "bad science", because it is impossible to check its usefulness for guiding action. This use of quantification will facilitate abuse and corruption. As noted by Porter (1995):

> *The appeal of numbers is especially compelling to bureaucratic officials who lack the mandate of a popular election, or divine right. Arbitrariness and bias are the most usual grounds upon which such officials are criticized. A decision made by the numbers (or by explicit rules of some other sort) has at least the appearance of being fair and impersonal. Scientific objectivity thus provides an answer to a moral demand for impartiality and fairness. Quantification is a way of making decisions without seeming to decide. Objectivity lends authority to officials who have very little of their own.*

## Section 3. The solution: establishing a new relation between science and governance

The paradigm of evidence based policy is based on the assumption of prediction and control, which may be used to eliminate "scruples", intended as feelings of doubt or hesitation with regard to the morality or propriety of a given course of action. Our suggestion is to take as a deliberate strategy the goal of reintroducing doubts and scruples in the process of deliberation, somewhat closer to Montaigne, somewhat farther from Descartes (Toulmin, 1990).

The task of guaranteeing the quality of the process of production and use of scientific information for governance must have as objective to minimize the negative effect of hypocognition on the final choice of a policy. For this reason it is essential to study the process of formalization of the chosen issue definition, e.g. how the frame was constructed, in semantic terms, and how this selection has cascaded into a predefined set of data, indicators and mathematical models. In this section we try advance a few suggestions to this end.

## 3.1 Responsible use of quantitative information

A first requirement for a better use of science for policy is a responsible use of quantitative information, away from indicators rich in spurious accuracy and fantastic model-generated numbers. This requires the adoption of specific tools of quality control. Practical tools developed in the context of PNS to address these topics are NUSAP and sensitivity auditing.

- NUSAP is a notational system called for the management and communication of uncertainty in science for policy, based on five categories for characterizing any quantitative statement: Numeral, Unit, Spread, Assessment and Pedigree (Funtowicz & Ravetz, 1990; van der Sluijs et al., 2005; see also http://www.nusap.net/).

- Sensitivity auditing (Saltelli et al., 2013; Saltelli and Funtowicz, 2014) extends sensitivity analysis as used in the context of mathematical modelling to settings where the models are used to produce inference for policy. Sensitivity auditing questions the broader implications of the modelling exercise, its frame, its assumptions, the assessment of the uncertainties, the transparency of the inference, the veracity of the sensitivity analysis and the legitimacy of the assessment.

## 3.2 Taming scientific hubris

Frank H. Knight observed in 1921 that:

> We live in a world of contradiction and paradox, a fact of which perhaps the most fundamental illustration is this: that the existence of a problem of knowledge depends on the future being different from the past, while the possibility of the solution of the problem depends on the future being like the past.

We suggest a re-learning of Knight's warning, and a stronger reconsideration of the differences between risks, that can be computed, versus uncertainties, which cannot. Ignoring this lesson will transform us in the character of the joke where a drunkard looks for his lost key under the lamppost, even though he knows that he lost it elsewhere, only because at least under the post there is light. Nassim Nicholas Taleb calls this 'The delusion of uncertainty'. With Bryan Wynne (1992) we would also like to extend this taxonomy of uncertainties to include dimensions of ignorance and indeterminacy, i.e.:

> RISK - *Know the odds.*
> UNCERTAINTY - *Don't know the odds: may know the main parameters. May reduce uncertainty but increase ignorance.*
> IGNORANCE - *Don't know what we don't know. Ignorance increases with increased commitments based on given knowledge.*
> INDETERMINACY - *Causal chains or networks open.*

For Wynne:

> *Science can define a risk, or uncertainties, only by artificially 'freezing' a surrounding context which may or may not be this way in real-life situations. The resultant knowledge is therefore conditional knowledge, depending on whether these pre-analytical assumptions might turn out to be valid. But this question is indeterminate - for example, will the high quality of maintenance, inspection, operation, etc., of a risky technology be sustained in future, multiplied over replications, possibly many all over the world?*

The taming of scientific hubris is at the basis of a more effective use of science for governance.

### 3.3 Evidence based policy versus robust policy

We suggest moving beyond 'evidence based policy' – and its related hyper quantification based on simplified narratives – toward 'robust policy' based on a strategy of filtering of potential policies in a context of falsification. We call this 'robust policy' borrowing from Helga Nowotny's (2003) concept of socially robust knowledge: a kind of knowledge that has been filtered through the lenses of different stakeholders and normative stances.

Thus the quality check on proposed policies and narratives about governance could be carried out using the method of falsification with respect to:

- feasibility (compatibility with external constraints),
- viability (compatibility with internal constraints) and
- desirability (compatibility with normative values adopted in the given society).

If the policy will result unfeasible or unviable or undesirable in relation to one of the quality checks we would have individuated either a bottleneck or a political issue or a true impossibility to be dealt with.  No more prediction and control leading to planning and optimization, but rather strategic learning through falsification leading to flexible management.

This approach has elements of similarity with the strategy suggested by Rayner (2012) to overcome socially constructed ignorance: the idea of 'clumsy solutions'. While socially constructed ignorance helps to keep 'uncomfortable knowledge' at bay, clumsy solutions allow it to be processed:

> *Clumsy solutions may emerge from complex processes of both explicit and implicit negotiation. In other words, solutions are clumsy when those implementing them converge on or accept a common course of action for different reasons or on the basis of unshared epistemological or ethical principles […] They are inherently satisficing […] rather than optimizing approaches, since each of the competing solutions is optimal from the standpoint of the proposer. Clumsy solutions are inherently pluralistic […]*

Clumsy solutions resonate with the 'working deliberatively within imperfections' of the Post Normal Science's extended participation model, and with the 'rediscovery of ignorance' advocated by Ravetz (2015, p. xviii).

A key step in the identification of the feasibility, viability and desirability domains entails looking at the state of affair through different lens – i.e. dimensions and scales of analysis  This predicament for quantitative analysis becomes systemic in sustainability analysis (Giampietro et al. 2013; 2014): when checking the feasibility of food security against external constraints (context/black-box - agriculture) we have to measure requirements and supply in terms of kg of potatoes, vegetables and animal products. However, if we want to check the viability of food security in relation to internal constraints (black box/internal parts – human diet) we have to measure requirements and supply in terms of kcal of carbohydrates, proteins and fats.  In the same way, when checking the feasibility of energy security against external constraints (context/black-box - "primary energy sources") we have to measure relevant physical quantities in terms of tons of coal, kinetic energy of falling water, cubic meters of natural gas, whereas if we want to check the viability in relation to internal constraints (black box/internal parts - "energy carriers") we have to measure relevant quantities in terms of kWh of electricity, MJ of fuels. This epistemological predicament is generated by the fact that different types of quantitative assessments are non-equivalent and cannot be compressed to a single indicator (Giampietro et al. 2006). Quantitative representations useful to study *feasibility* are not equivalent to quantitative representations useful to study *viability* and the information given by these two typologies of representations cannot be used to study *desirability* without involving in the discussion those social actors carrying legitimate but contrasting normative values.

Only after having operationalized the definition of these three domains it becomes possible to carry out an informed deliberation for evaluating policies having the goal of balancing efficiency with adaptability in view of sustainability. This may possibly feed into a multi-criteria characterization of the proposed solutions with respect to the different normative ingredients.

The proposed approach is equivalent to exploring a multi-dimensional space with a parsimonious and appropriate experimental design, instead of concentrating an unrealistic degree of detail around a single point in this space.

### 3.4 Quantitative story telling for governance

The widening of the set of available 'frames' may be achieved via a quantitative 'story-telling for governance', with the goal of generating plausible and relevant stories capable of reducing hypocognition in the chosen issue definition/problem structuring – a strategy also suggested by 'cognitive activist' George Lakoff (2014).

Quantitative story telling has the goal of guaranteeing the quality of the chosen story-telling in the given socio-economic and ecological context. The quality of the story-telling has to do with the reduction of the negative effects of the hypocognition associated with the chosen problem structuring – the unavoidable neglecting of relevant narratives (relevant "known knowns" and "known unknowns") and the unavoidable presence of "unknown unknowns". The fitness of different policy options can then be gauged from the integration of a robust mix of relevant narratives, plausible explanations and pertinent perceptions.

This qualitative check of the coherence of the quantitative information generated by non-equivalent models is essential. In fact, models are by-products of the pre-analytical choice of relevant causal relations and data are by-products of the pre-analytical choice of relevant perceptions. Confronted with numbers coming from several non-equivalent descriptive domains (logically incoherent quantitative representations) one can no longer rely on big data and sophisticated algorithms. Without a quality check on the chosen story-telling more data and larger models developed within arbitrarily constrained explanations and perceptions will only increase the level of indeterminacy and uncertainty of the results.

The usefulness of the chosen stories needs to be validated using quantitative analysis that must remain coherent across scales and dimensions – i.e. a multi-scale integrated analysis of the functioning of socio-ecological systems, inclusive of their level of openness, e.g. to trade relationships, lest relevant aspects of the problems are simply externalized.

### 3.5 Getting the right narratives before crunching numbers

In his 'plea for reasonableness versus rationality' Stephen Toulmin (1990, 2001) contrasts the ideal of Renaissance Humanism against the Renaissance scientific revolution, which he considers as a counter-Renaissance, where Descartes' certainties replace the doubts of Montaigne. In order to return to reason, he warns, we need to 'do the right sums' more than we need to 'do the sums right' (2001, p.66). This implies a careful selection of the stories to be told before indicators are built, data collected and models run. We need to explore more frames as opposed to selecting just one and filling it with numbers.

We can illustrate this with the persisting controversy surrounding the use of genetically modified organism, a quintessential wicked issue.

The journal 'The Economist', discussing a GMO labelling scheme in Vermont (US) commented recently (2014):

> *Montpelier is America's only McDonald's-free state capital. A fitting place, then, for a law designed to satisfy the unfounded fears of foodies* […] *genetically modified crops, declared safe by the scientific establishment, but reviled as Frankenfoods by the Subarus-and-sandals set.*

For those unfamiliar with this kind of jargon, Frankenfood is GMO based food as defined by its opponents, while the *Subarus-and-sandals* set is an The Economist's own synecdoche to allude to those in Vermont who support a labelling scheme for GMO-containing food. The image accompanying the piece shows a hippy-looking public protesting against GMO. We use this as a vivid illustration of what any reader knows: opposition to GMO food is normally portrayed as a Luddite, anti-science position, and this because GMOs are treated as a nutritional 'risk to health issue'.  Against this irrational position science has 'spoken' by declaring GMO's safe for human consumption, thus modern societies should by law permit (or even force, in the name of progress) their production and consumption.

This frame clashes against the reality of citizens' true concern, as measured e.g.by Marris et al., (2001).  In the list of citizens' concerns gathered through participatory processes, the issue of food safety seems to be conspicuously absent, whereas a complete different set of crucial questions are instead posed:

- *Why do we need GMOs? What are the benefits?*
- *Who will benefit from their use?*
- *Who decided that they should be developed and how?*
- *Why were we not better informed about their use in our food, before their arrival on the market?*
- *Why are we not given an effective choice about whether or not to buy and consume these products?*

- *Do regulatory authorities have sufficient powers and resources to effectively counter-balance large companies who wish to develop these products?*

The variety of frames revealed by these concerns reveals that the prevailing frame 'safe GMO food versus recalcitrant citizens' is "irrelevant" for the decision to be taken.

## Conclusions

The evidence based policy paradigm should be revised because it uses science to reinforce hypocognition. Evidence based policy cannot be separated by policy based evidence. The accumulation of data, indicators and mathematical modelling in support to a given frozen framing of an issue obfuscates and distracts from the important task which is the semantic opening of the frame. A quantitative problem structuring may empower those that have selected the given story-telling to eliminate, through induced hypocognition, uncomfortable knowledge (Rayner, 2012). Spurious precision and disproportionate mathematics deter the use of plain English to question the premises of an analysis. This is not a new finding. Already in 1986 (p.138-154) Langdon Winner warned ecologists not to fall into the trap of risk and cost benefit analyses:

> *"[T]he risk debate is one that certain kinds of social interests can expect to lose by the very act of entering. (…) Fortunately, many issues talked about as risks can be legitimately described in other ways. Confronted with any cases of past, present, or obvious future harm, it is possible to discuss that harm directly without pretending that you are playing craps. A toxic waste disposal site placed in your neighborhood need not be defined as a risk; it might appropriately be defined as a problem of toxic waste. Air polluted by automobiles and industrial smokestacks need not be defined as a 'risk'; it might still be called by the old-fashioned name, 'pollution'. New Englanders who find acid rain falling on them are under no obligation to begin analyzing the 'risks of acid rain'; they might retain some Yankee stubbornness and confound the experts by talking about 'that destructive acid rain' and what's to be done about it. A treasured natural environment endangered by industrial activity need not be regarded as something at 'risk'; one might regard it more positively as an entity that ought to be preserved in its own right".*

Complex adaptive systems, be these the labour market of a country or its forests, are reflexive and continuously becoming "something else" in order to reproduce themselves (Prigogine, 1980). For this reason it is impossible to predict deterministically their future states, because if they manage to reproduce themselves in the long run, they can do so only by moving in a finite time between states that have to be simultaneously:

- feasible – compatible with boundary conditions determined by process outside human control;
- viable – compatible with the structure of internal parts and their system of control;
- desirable – compatible with normative values used to legitimize the social contract keeping together the social fabric.

In this situation any definition of what should be considered as "feasible", "viable" and "desirable" has to be continuously updated. Changes in the processes outside human control will determine changes in boundary conditions (changing the definition of the feasibility domain), in the same way as the full range of consequences of changes taking place in processes under human control, when spreading across scales and dimensions, may impact both the viability and the desirability of a trajectory, again impairing long term deterministic predictions.

In human societies, shared normative values (frames) are the result of negotiation and shifting of power relations (Lakoff, 2014). This implies that even those story-tellings, strategies, and narratives that resulted useful for guiding human action in a given historic period may become useless (and therefore potentially dangerous as misleading) when the meanings they assign to the terms "feasibility", "viability" and "desirability" in relation to the stated goals has changed. To make an example, the 'Endless Frontier' metaphor is today less convincing than it was right after Word War II. Not only is now science perceived as an instrument of profit and power as discussed in section 1, but its crisis has dramatically curtailed its rate of progress (Le Fanu, 2009, 2010). In this respect the acritical adoption of prevailing narratives can be fatal. Already in the XIX century Giacomo Leopardi considered 'Fashion' more deadly that 'Death'.

To make another example at the hearth of present economic dispute the neo-classical Economics narrative of perpetual growth based on continuous innovation supposed to reduce inequity through a trickle-down effect was meaningful for developed economies experiencing a period of maximum expansion in their pace of economic activity. As suggested by Daly (1992) the evolution from an "empty world" (a low population on the planet) to a "full world" (a large population on the planet) means more stringent external constraints and a larger environmental impact of human activity. This translates into a reduction in the pace of economic expansion and an increase in inequality within societies – i.e. an increase in troubles with internal constraints. As an example of what is meant by trouble one can recall here the present debate on the virtues and faults of capitalism *qua* capitalism (Piketty, 2013; Bellamy Foster and Yates, 2014), and the linkages between increased inequality and increasing scope for rent seeking and corruption by the elites (Acemoglu and Robinson, 2012).

As any re-adjustment of normative values (requiring large power shifts) is problematic, what is the role the science should play in this re-adjustment of the meanings to be assigned to the concept of "feasibility", "viability" and "desirability"?

So far, the role of science for governance and sustainability has been that of a driver of techno-scientifical of human progress. As discussed in section 2 this strategy represents a full endorsement of the ideas of Bacon, Condorcet and Bush that all human problems can be solved by science and technology. Many institutional actors seem to be mired in the old strategy of prediction and control, making plans informed e.g. by the results of Dynamic Stochastic General Equilibrium models whose inadequacy we discussed in Section 2, or by cost benefit analyses of the impact of climate on economy and society extending one hundred year into the future (Saltelli and d'Hombres, 2011, Rhodium Group, 2014). A disturbing symptom of this practice is the delirious precision of the estimates we are fed with.

* '*D.C. climate will shift in 2047*' (Bernstein, 2013);

* '*August 22 was Earth Overshoot Day. In 8 Months, Humanity Exhausted Earth's Budget for the Year*' (Global Footprint Network, 2014); for a criticism of this estimate see Giampietro and Saltelli (2014).

Used in this distorted way mathematical modelling finishes to play the religious or ritualistic role of haruspices. The use of the term 'ritual' is not exaggerated here. Nobel laureate Kenneth Arrow experienced the same impression. During the Second World War he was a weather officer in the US Army Air Corps working on the production of month-ahead weather forecasts, and this is how he tells the story (Szenberg, 1992):

> *The statisticians among us subjected these forecasts to verification and they differed in no way from chance. The forecasters themselves were convinced and requested that the forecasts be discontinued. The reply read approximately like this: "The commanding general is well aware that the forecasts are no good. However, he needs them for planning purposes".*

Richard Feynman in his Caltech's 1974 Commencement Address "Some remarks on science pseudoscience and learning how not to fool yourself" calls this type of use of science as "cargo cult science":

> "*In the South Seas there is a cargo cult of people. During the war they saw airplanes land with lots of good materials, and they want the same thing to happen now. So they've arranged to imitate things like runways, to put fires along the sides of the runways, to make a wooden hut for a man to sit in, with two wooden pieces on his head like headphones and bars of bamboo sticking out like antennas--he's the controller--and they wait for the airplanes to land.*"
> "*They're doing everything right. The form is perfect. It looks exactly the way it looked before. But it doesn't work. No airplanes land. So I call these things cargo cult science, because they follow all the apparent precepts and forms of* *scientific investigation, but they're missing something essential, because the planes don't land.*"

Science provided to western societies incredible achievements, and many believe that the same achievements will be now obtained also by all the other countries. What is missed by those holding this hope is the awareness that many problems solved by progress in western countries were solved by externalizing to someone else (the environment, future generations, other countries) the negative consequences of the increased level of consumption of resources per capita. Things are quite different when considering the sustainability of technical progress of the whole world. At the global level there is no room for externalization – it is a zero-sum game, what goes around comes around – and there is no "free-lunch" - someone is paying or will pay what is consumed. We must thus deal with an inextricable nexus between physical, biological, social and ethical issues. The hope that this issue will be tackled by more computer power, more complicated models, huge databases, and more rigour in the scientific method leaves us with a "sustainability science" which resembles "cargo cult science". The problem with this type of sustainability science is that, as suggested by Feynman, in spite of the frenetic activity of its practitioners "planes don't land".

---

[i] A very concise summary of what 'science wars' means is in Sarewitz, 2000. Wikipedia's entries for 'science wars' and 'two cultures' are also informative. The Canadian Broadcasting Corporation CBC has an excellent series 'How To Think About Science' with an interview with historian of science Simon Schaffer, see
http://www.cbc.ca/ideas/episodes/2009/01/02/how-to-think-about-science-part-1---24-listen/